\documentclass[aps,prl,twocolumn,groupedaddress]{revtex4}
\usepackage{graphicx}
\usepackage{amsmath}
\usepackage{amsfonts}
\usepackage{amssymb}
\usepackage{epsfig}
\usepackage{color}
\usepackage{bm}

\newcommand{\fii}{\varphi}
\newcommand{\e}{\varepsilon}
\begin{document}

\title{Ion energy scaling under optimum conditions of laser plasma acceleration}

\author{A. V. Brantov}
\affiliation{P. N. Lebedev Physics Institute, Russian Academy of
Science, Leninskii Prospect 53, Moscow 119991, Russia}
\author{E. A. Govras} \affiliation{P. N. Lebedev Physics Institute,
Russian Academy of Science, Leninskii Prospect 53, Moscow 119991,
Russia}
\author{V. Yu. Bychenkov} \affiliation{P. N. Lebedev Physics Institute,
Russian Academy of Science, Leninskii Prospect 53, Moscow 119991,
Russia}
\author{W. Rozmus}
\affiliation{Theoretical Physics Institute, University of Alberta,
Edmonton T6G 2E1, Alberta, Canada}

\begin{abstract}
A new, maximum proton energy, $\e$, scaling law with the laser pulse energy, $E_L$
has been derived from the results of 3D particle-in-cell (PIC) simulations. Utilizing numerical modelling, protons are accelerated during interactions of the femtosecond relativistic laser pulses with the plain semi-transparent targets of optimum thickness [Esirkepov {\it et al.} Phys. Rev. Lett. {\bf 96}, 105001 (2206)]. The scaling, $\e \sim E_L^{0.7} $, has been obtained
for the wide range of laser energies, different spot sizes, and laser pulse
durations. Our results show that the proper selection of foil target optimum thicknesses, results in a very promising increase of the ion energy with the laser intensity even in the
range of parameters below the radiation pressure (light sail) regime. The proposed analytical model is consistent with numerical simulations.

\end{abstract}
\maketitle
Ion acceleration by intense ultra-short laser pulses has led to many original
application such as: triggering of nuclear reactions \cite{bychenkov_99,labaune_13};
research of warm dense matter \cite{warm_dense_matter}; laboratory astrophysics
\cite{astrophysics}; radiography \cite{radiography_1,radiography_2,willingale_10}; fast ignition \cite{roth_01};
and, hadron therapy \cite{bulanov_02}. Both experiments \cite{exp_1,exp_2} and simulations \cite{sentoku_02} have demonstrated an increase of ion energy with a corresponding reduction of the target thickness. The high contrast pulses of modern lasers \cite{exp_1,exp_2} have enabled effective acceleration of ions from ultra-thin
foils that are semi-transparent to laser light. In this regime, a high-intensity
laser pulse expels electrons from the irradiated area of the foil in a forward
direction that causes ion acceleration from the entire target volume through the
mechanism combining elements of the Target Normal Sheath Acceleration (TNSA) mechanism and the Coulomb explosion, i.e. the directed Coulomb explosion \cite{bulanov_08} or alternatively interpreted as the "break-out afterburner"
\cite{BOA}.

It is known that target thickness should be properly matched to the laser
intensity \cite{esirkepov06} in order to obtain maximum ion energy. Although the optimum
target thickness can be estimated in the order of magnitude from 2D PIC
simulations \cite{esirkepov06}, the 3D PIC simulations are needed to correctly
quantify target thickness for different laser intensities.  The optimum target thickness was found in Ref. \cite{esirkepov06}, wherein the square root energy scaling for the proton energy with laser intensity (energy) was also inferred. The square root intensity scaling for the maximum proton energy is also predicted by the TNSA model,
which is valid for thick targets \cite{fuchs06}. The 3D PIC simulations to find optimum target thickness for the laser pulse energy must also examine the dependence on different spot sizes and pulse durations. Such
optimization should include a systematic study of laser light absorption
(i.e. laser to electron energy conversion efficiency) in semi-transparent
targets: a study that will form an important part of our paper. Our comprehensive study will quantify how all
parameters of the pulse affect laser energy conversion to hot electrons and define
the effectiveness of high-energy ion production. We will present the results of our 3D
optimization study with PIC code Mandor \cite{mandor} for acceleration of protons from thin
targets triggered by femtosecond laser pulses. The main outcome of this
paper is the dependence of maximum proton energies on the laser intensities under conditions of volumetric heating of optimum thickness targets.

We also propose a simple theoretical model that estimates maximum ion
energy using the effective temperature of the laser heated electrons. Since
the electric field that accelerates ions depends on the charge separation, the ratio of the Debye length of laser heated electrons $\lambda_{D_e}$ to the foil thickness $l$ is the main controlling parameter of
the theory. So far, only two asymptotic cases have been studied in detail:
quasineutral expansion $\lambda_{D_e}\ll l$ \cite{Mora_PRL_2003} and Coulomb
explosion $\lambda_{D_e}\gg l$ \cite{Kovalev_QE_2005}. None of these theories are fully applicable to thin semitransparent targets. We will show that our model, that is valid
for arbitrary $\lambda_{D_e}/ l$, qualitatively explains numerical simulations
when the ponderomotive dependence \cite{wilks01} of the effective electron temperature on the
absorbed laser energy is used.

It has been already shown that thin foils with optimum thicknesses are much better for
proton acceleration. Making use of shorter laser pulses or tight laser focusing
also results in some increase of proton energies \cite{humieres13}. Building on these results, we have performed 3D simulations of proton acceleration by ultrashort ($\tau = 30$ fs FWHM duration) tightly focused
(FWHM of the laser spot $d = 4 \lambda$) laser pulses with the PIC code MANDOR
\cite{mandor}. As a reference we set the laser wavelength at $\lambda = 1 \mu$m. Laser intensity is varied from $I=5 \times 10^{18} $ to $I=5 \times 10^{22}$ W/cm$^2$, which corresponds to laser pulse energies from
0.03 J to 300 J. To examine pulse duration ($\tau$) and spot
size ($d$) effects on ion acceleration, three additional sets of laser
pulse parameters with the same full laser energy have been used: increased laser
spot size, $d = 6 \lambda$, and decreased one,
$d = 2 \lambda$, for $\tau = 30$ fs and for the increased pulse duration, $\tau = 150$ fs, for the hot spot size $d = 4
\lambda$. A very tight laser pulse focusing to $2 \lambda$ has
been implemented by using the parabolic mirror simulation technique
\cite{popov}. For larger hot spots, Gaussian laser pulses have been
used.

The laser pulse was focused on the front side of the thin CH$_2$ plasma foils, that are composed
of electrons, hydrogen ions, and fully ionized carbon ions ($C^{6+}$). The target has an electron density of $200
n_{c}$ (where $n_{c} = m_e/4 \pi e^2 \omega^2$ is the plasma critical
density), which corresponds to a solid mass density of CH$_2$ ($1.1$
g/cm$^3$). The target thickness $l$ has been varied from $3$ nm
to $1$ $\mu$m.

\begin{figure} [!ht]
\centering{\includegraphics[width=7.2cm]{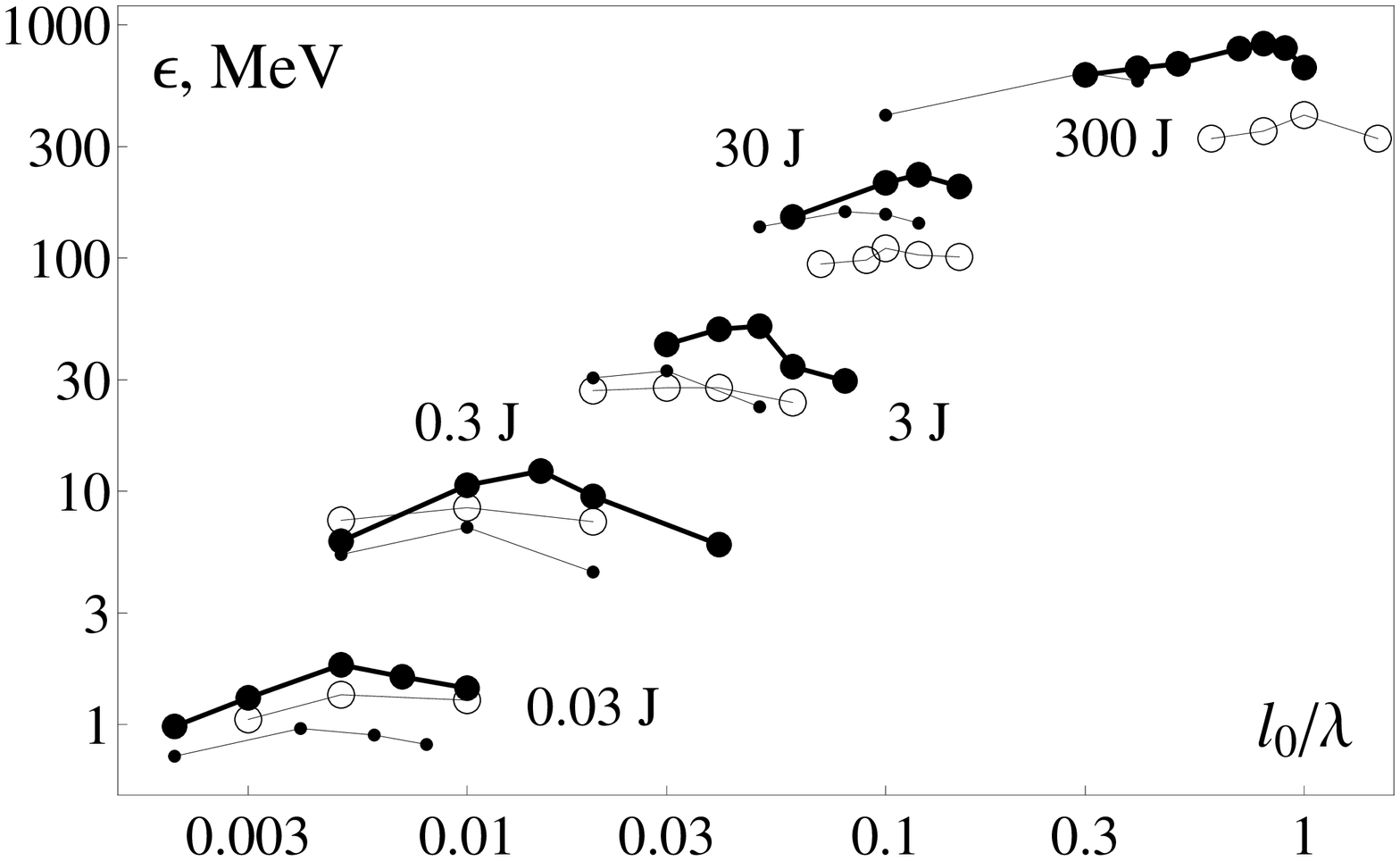} \hspace{0.6cm}
\includegraphics[width=7.2cm]{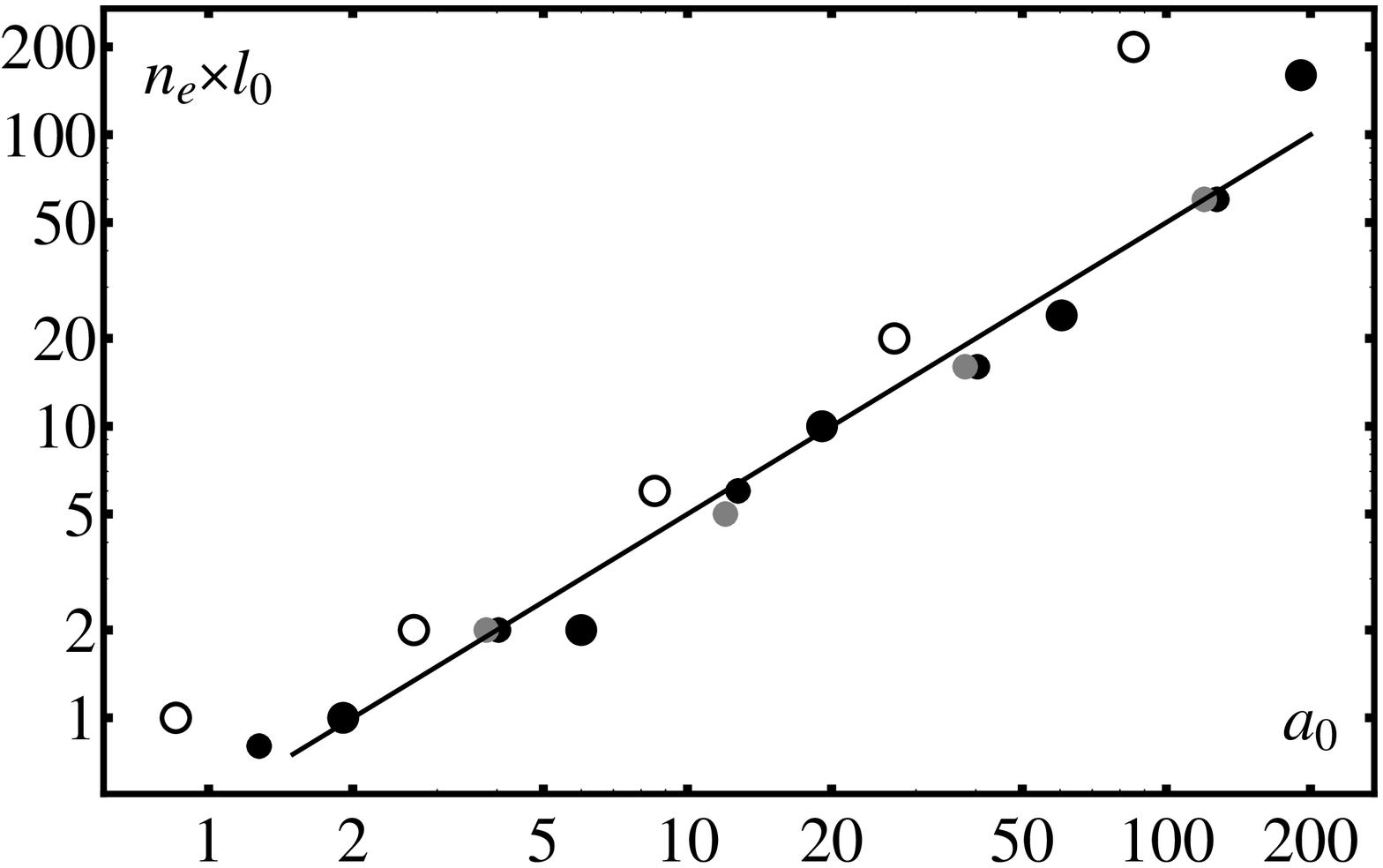}} \caption{Maximum proton
energy vs. target thickness for different laser energies (top panel) and optimum target areal density in $n_c \times \lambda$ units vs. laser field $a_0$ (bottom panel). For both panels large black dots stand for $\tau = 30 fs$ and $d=4 \mu$m,
small black dots for $\tau = 30 fs$ and $d=6 \mu$m, gray dots for $\tau = 30 fs$
and $d=2 \mu$m, open circles for $\tau=150 fs$ and $d=4 \mu$m.}\label{fig1}
\end{figure}

We performed several runs with different target
thicknesses that have been chosen according to the theoretically predicted optimum length, $\lambda\,a_0\,n_c/ (n_e \pi)$ \cite{vshivkov}, where
$a_0=0.85\sqrt{I[\text{W/cm}^2]\,\lambda[\mu
\text{m}]^2\,10^{-18}\vphantom{\int}}$. The results of the
simulations for maximum proton energies are shown in the top panel of Fig.~\ref{fig1}. They clearly
confirm the existence of an optimum target thickness for which protons reach maximum energies.
The optimum target thickness corresponding to the maximum of ion energies (Fig. \ref{fig1}, bottom panel) grows linearly with laser field amplitude, $l_0 =  0.5 \lambda a_0 n_c/n_e$, where the numeric factor 0.5 is almost
independent from the focal spot size and it is slightly above the theoretically
predicted 1/$\pi$ \cite{vshivkov}. At the same time,
this factor slightly increases with the laser pulse duration (see open circles
in Fig. \ref{fig1}, bottom panel), i.e. the optimum regime of ion acceleration
by the longer pulses requires thicker targets.
The optimum target thickness corresponds to the partially
transparent target \cite{vshivkov}, when volumetrically heated
electrons are swept out of the plasma in a forward direction and give rise to strong
charge separation fields \cite{brantov13}, which accelerate ions by the directed Coulomb
explosion \cite{bulanov_08}. The laser pulse can then expel a large number of hot electrons from the hot spot region and penetrate inside the target in a hole-boring like action. This is the reason why optimal target thickness  is above theoretically predicted values and this is why longer laser pulses can penetrate through thicker targets.

\begin{figure} [!ht]
\centering{\includegraphics[width=8.2cm]{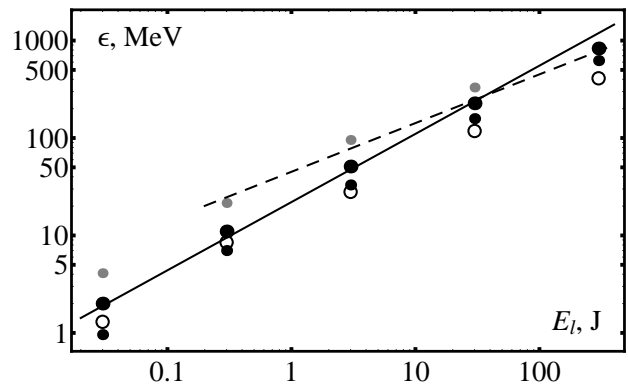}} \caption{Maximum
proton energy for optimum target thickness vs. laser energy
for $\tau = 30 fs$ and $d=4 \mu$m (large black dots), $d=6 \mu$m (small black
dots), $d=2 \mu$m (gray dots) and $\tau=150 fs$ and $d=4 \mu$m (open
circles).
The black and dashed lines correspond to the fits $\e[\rm{MeV}]= 22 (E_L[\rm{J}])^{0.7}$ and
$\e[\rm{MeV}]= 45 (E_L[\rm{J}])^{0.5}$ \cite{esirkepov06}, correspondingly.}\label{fig2}
\end{figure}

Figure \ref{fig2} shows the dependence of ion maximum energy on the laser pulse energy.
The results of our simulations for the maximum energy of the protons from the optimum thickness targets are well approximated by the scaling $\e \propto E_L^{0.7}$, which is different from the square root dependence reported earlier \cite{esirkepov06}. The numerical proportionality factor varies from $14$ for a long pulse (150 fs) and large focal spot diameter ($d\gtrsim 4\lambda$), to $45$ for tight laser focusing into the focal spot of  d=$2 \lambda$. Given laser energies the shorter laser pulse and tighter focusing gives higher maximum proton energy. The number of energetic protons also increases with the laser energy. Note that long pulses are more effective for proton acceleration if the laser energy is less than 1 J.

\begin{figure} [!ht]
\centering{\includegraphics[width=7.2cm]{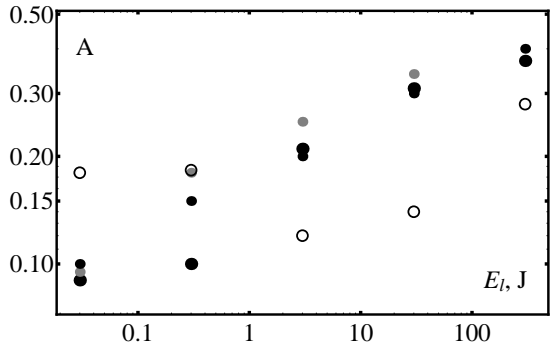}}
\caption{Laser light absorption in the targets of optimum thickness as function of the laser energy. Definitions of the dots are the same as in Fig. \ref{fig2}
}\label{fig3}
\end{figure}

To better understand the proton energy scaling of Fig. \ref{fig2}, we have analyzed the laser energy absorption coefficient, $A$, for semi-transparent targets. We define an absorption coefficient as the ratio of the particle kinetic energy to the initial laser energy. We found that for our parameters, the absorption coefficient of 30 fs pulse in the targets with optimum thickness increases with laser energy (see fig.~\ref{fig3}) from values less than 10\% for 0.03 J laser to 30 \% for 30 J laser.
The absorption of longer laser pulses is better for small laser energies. It is also not a monotonic function of the pulse energy. It drops from 18\% to 12\% at 3 J laser pulse energy and then grows to $\sim 30 \%$ for 300 J laser.

We will now demonstrate that the simple semi-analytic theory qualitatively agrees with our
simulation results and captures the main elements of interaction physics and ion acceleration. Assume that ion plasma occupies the layer of the width $l$ along
$x$ axis. The transversal size of the plasma is limited to the laser focal spot area, $\pi d^{\,2}/4$. For simplicity, we consider one ion species plasma where ions are initially at rest within $-l/2<x<l/2$. During
subsequent ion outflow, electrons remain in equilibrium with the electrostatic
field and are described by the Boltzmann distribution function with an effective temperature $T$. The
latter depends on laser intensity and can be time dependent due to adiabatic cooling after the pulse is terminated.
Plasma expansion at $t>0$ will be symmetrical with respect to the plane $x=0$. For small distances
$x_f<d$, where $x_f(t)$ is the ion front position, three-dimensional effects
due to transversal size of the plasma region can be neglected and outflow
is treated as one-dimensional. Motion of ion plasma will be described by the
following system of equations:
\begin{equation}
 \label{eq_system}
 \begin{aligned}
  &\fii''=\eta_e e^{\fii/T}-n_i(x,t),\\
  &\left.\fii'\right|_{x=0}=0,\;\left.\fii'\right|_{x=\infty}=0,\\
  &\ddot{x}=-\frac{\partial\,\fii}{\partial\,x},\;\dot{x}(0)=0\\
  &x(0)=x_0,\; 0\leq x_0\leq 1,\\
  &n_i(x,t)=\left|\partial x/\partial x_0\right|^{-1}
 \end{aligned}
\end{equation}
where $\eta_e=n_{e_0}/Z\,n_0$ is defined by the initial electron density in
the foil center $n_{e_0}$, initial ion density $n_0$, and
charge $Z$. The parameter $\eta_e$ has a simple approximate dependence on the initial electron
temperature $\eta_e=\left[1+2\,T(0)\vphantom{\int}\right]^{-1}$
\cite{Govras_JETPL_2013}. We have used the following
dimensionless variables in the system of equations \eqref{eq_system}: spatial coordinate $x$ is
measured in the unit of $l/2$, time $t$ is normalized to $1/\omega_{pi}$  and energies $Z\,T$ and  $Z\,e\,\fii$ are written in terms of $4\,\pi(Z\,e)^2\,n_0\,(l/2)^2$. Thus far there are no  analytic solutions for
Eq. \eqref{eq_system} at arbitrary $T$. The only known solutions are for the case
$\lambda_{D_e}\ll l$, i.e. for the quasineutral expansion, $T\to 0$,
\cite{Mora_PRL_2003} and for the case of $\lambda_{D_e}\gg l$, that is for the Coulomb explosion,
$T\to \infty$ \cite{Kovalev_QE_2005}. For the first case of the quasineutral expansion $Z n_i\approx n_e$ and
electric field at the ion front is $E_1=2\sqrt{T}/\sqrt{2\,e+t^2}$ \cite{Mora_PRL_2003}, where $e$=2.71828.... In the second limit of a Coulomb explosion, one finds $n_i=1/x_f(t)$ and $E_1\equiv 1$ where $x_f(t)$ is the front position of the expanding
ion plasma \cite{Kovalev_QE_2005}.

To find an approximate solution of  Eqns. \eqref{eq_system} for the arbitrary ratio of $\lambda_{D_e}/l$
we introduce the interpolating expression for $n_i(x,t)$ that is valid for an arbitrary temperature $T$. From the Poisson equation in Eq. \eqref{eq_system} one obtains $E_1=\int_0^{x_f}\left(n_i(x,t)-\eta_e\exp[\fii(x,t)/T]\right)dx$ where we used $n_e(x,t)=\eta_e\exp\left[\fii(x,t)/T\vphantom{\int}\right]$. If we choose
$n_i(x,t)=\eta_e\,\exp[\fii/T]+E_1/x_f$ the expression for $E_1$ is satisfied
and one obtains a correct asymptotics for both quasineutral outflow and for the
Coulomb explosion regime.
Solving Poisson equation in \eqref{eq_system} with such an ion density, we obtain
an implicit expression for the function $E_1(x_f)$:
\begin{equation}
 \label{E1}
 (E_{1})^{-1} = 1+\sqrt{\frac{x_f\,E_{1}}{2\,T}}
\exp\left[\frac{x_fE_{1}}{2\,T}\right]\,Erf\left[\sqrt{\frac{x_f E_{1}}{2\,T}}\right].
\end{equation}
Here $Erf(z)=\int_0^ze^{-t^2}dt$. Note that Eq. \eqref{E1} is simpler than that
in Ref.\cite{Govras_JETPL_2013} and has no interpolation coefficient.

We assume that plasma electrons are heated during the laser pulse duration $t<\tau$ and reach the characteristic
temperature $T_0$. However, after that, $t>\tau$, they adiabatically cool down as
described in \cite{Kovalev_JETP_2002}. Temporal behaviour of electron
temperature can be described as follows \cite{Kovalev_JETP_2002}
\begin{equation}
 \label{eq_T_t}
 T(t)=T_0\,\left[\Theta(\tau-t)+\frac{\Theta(t-\tau)}{1+(t-\tau)^2/t_c^2}\right],
\end{equation}
where $\Theta(t)$ is the Heaviside step function and the characteristic cooling time
is defined as $t_c=L/\sqrt{2}\,c_s$. Where $L$ is the spatial scale of ion
density and $c_s$ is the ion sound speed. We chose $L=x_f(\tau)$ and
$c_s=\sqrt{T_0}$ as the typical parameters.

When ions reach the distance $x_f\sim L_1= 1+d$ three-dimensional effects must be taken into
account. For $x_f\gg L_1$ electric field decreases $\propto x^{-2}$. Ensuring that the electric field satisfies two limits at the ion front, $E_1$ \eqref{E1} for $x_f<L_1$ and $E_1(L_1)/(x-L_1)^2$
for $x_f\gg L_1$ one may propose a smooth connecting expression valid for the front position $x_f$ at the arbitrary distance from the foil. In addition, a laser pulse introduces asymmetry into plasma expansion because all electrons are accelerated by the laser pulse in a forward direction on the rear side of the thin foil. We will assume that the electric field at $x>0$ is twice the value $E_1$ \eqref{E1} defined above for the symmetric expansion of the hot plasma layer into a vacuum.
Consequently, the electric field at the position of the ion front at the arbitrary time can be approximately
written as follows:
\begin{equation}
 \label{eq_E}
 E(x_f)=\left\{
 \begin{aligned}
  &2\,E_1(x_f), x_f\leq L_1,\\
  &2\,E_1(L_1)\left[1+(x_f-L_1)^2\vphantom{\sqrt{A}}\right]^{-1}, x_f>L_1,
 \end{aligned}
 \right.
\end{equation}
where the time evolution of the electron temperature that contributes to $E_1$  \eqref{E1} is given by the Eq.
\eqref{eq_T_t}. Solving the equation of motion \eqref{eq_system} for the ions at the front of the expanding plasma  with the electric field $E(x_f)$ \eqref{eq_E} one obtains maximum ion energy $\e_{max}=(\dot{x}_f)^2/2$.

\begin{figure} [!ht]
\includegraphics[width=7.2cm]{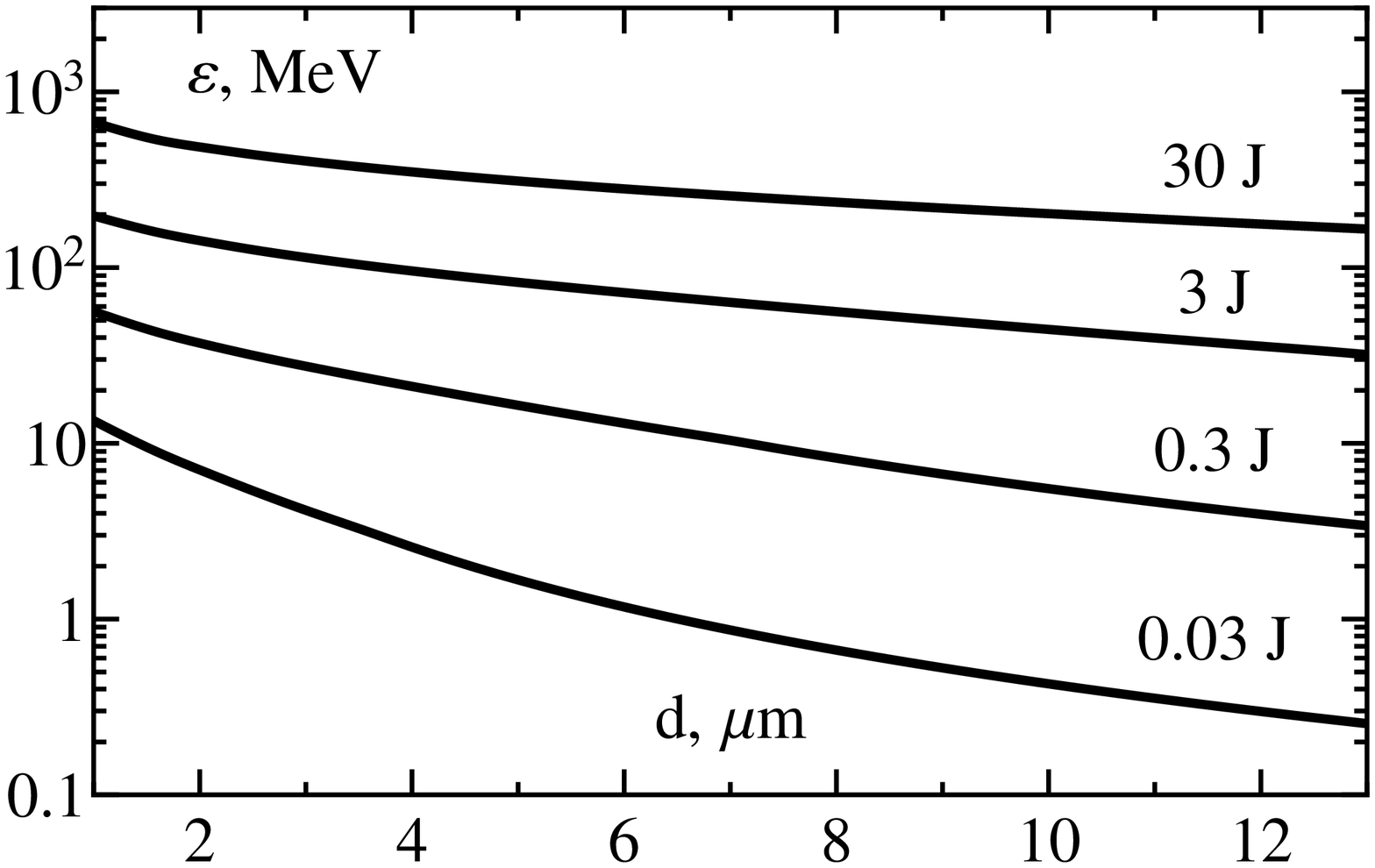}
\hspace{0.2cm}\includegraphics[width=7.5cm]{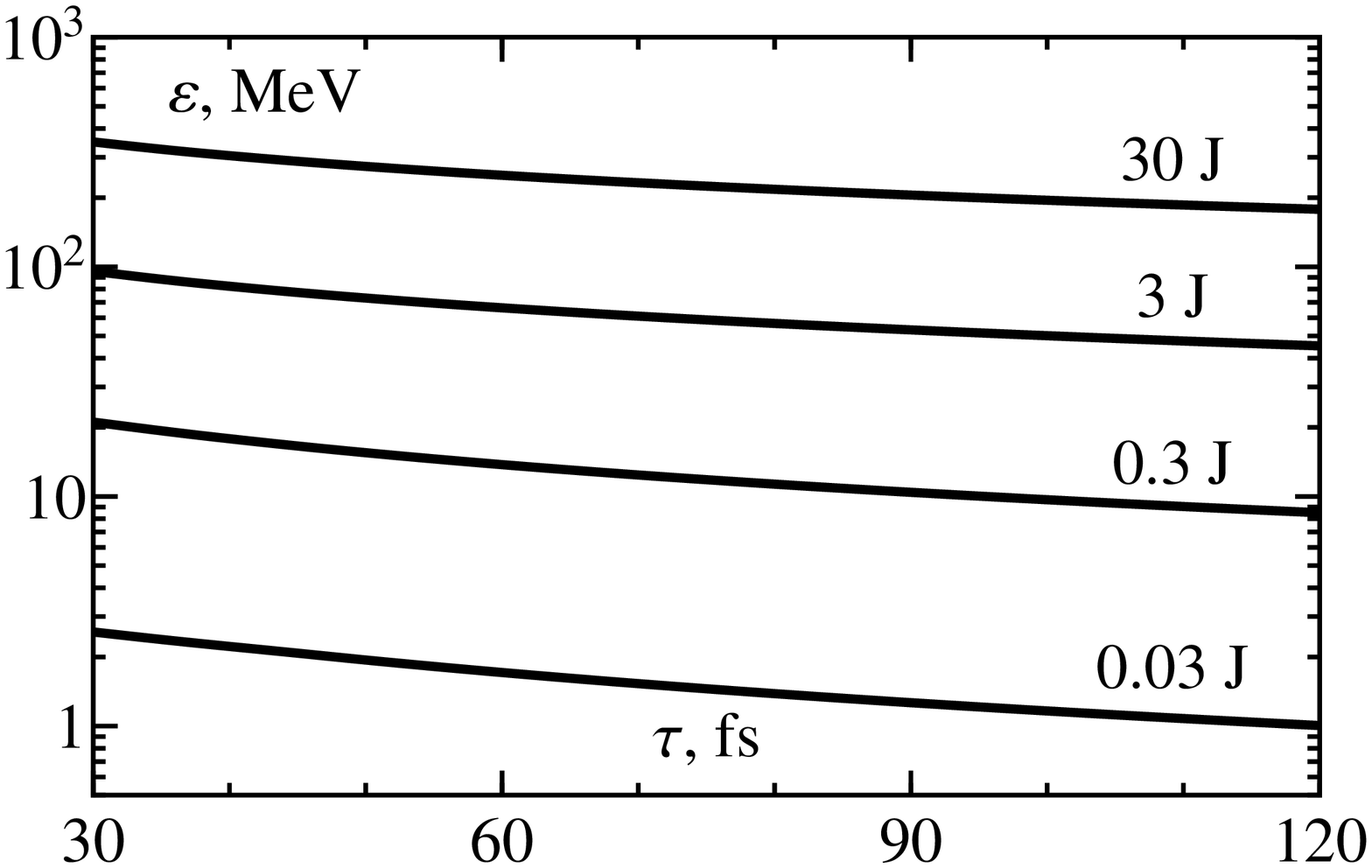}
\caption{Proton maximum energy vs. hot spot size $d$ (top panel) and vs. pulse duration (bottom
panel) for a hydrogen foil of $n_e=200\, n_{cr}$ and optimal thickness.}\label{pic_Df_tau}
\end{figure}

To illustrate the results of our theoretical model we have plotted the maximum ion energy for the
hydrogen foil as a function of the laser spot diameter (Fig. \ref{pic_Df_tau}, top panel) and the pulse duration
(Fig. \ref{pic_Df_tau}, bottom panel).
We have assumed that the electron temperature $T_0$ obeys the ponderomotive scaling law \cite{wilks01}: $T_0
\propto m\,c^2(\sqrt{1+(a_1)^2\vphantom{\int}}-1)$, where $a_1$ is
the laser amplitude, calculated using the absorbed fraction of laser intensity $a_1=a_0\sqrt{A}$. The foil
is assumed to be of an optimal thickness $l_0$. Plots in Fig.
\ref{pic_Df_tau} agree with simulations (c.f. Fig. \ref{fig2}) to the extent that for a given pulse energy, a defocussed laser light as well as a longer pulse duration lead to a maximum ion energy decrease.

\begin{figure} [!ht]
\centering{\includegraphics[width=8.2cm]{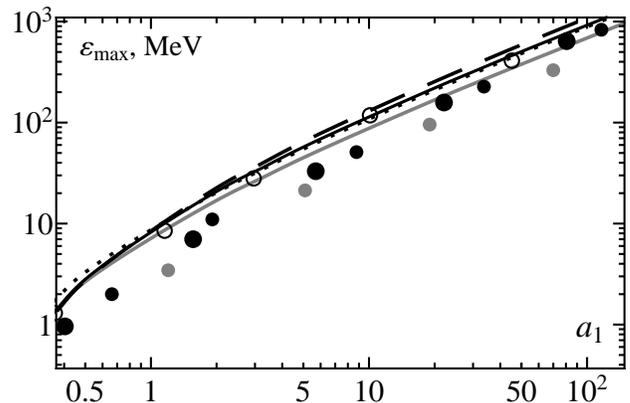}} \caption{Maximum
proton energy vs.
absorbed laser field amplitude $a_1$. Circles represent the numerical simulations and
lines are the theoretical results for the
following laser parameters:
$d = 2\mu$m, $\tau = 30$ fs (gray dots and solid gray line), $d = 4\mu$m,
$\tau = 30$ fs (small black dots and solid black line), $d = 6\mu$m,
$\tau = 30$ fs (large black dots and dashed black line), $d = 4\mu$m,
$\tau = 150$ fs (open circles and dotted black line).}\label{pic_e_a1}
% Definitions of the dots is the same as in Fig. \ref{fig2}.}
\end{figure}

In Fig. \ref{pic_e_a1} we compare
maximum energy of the protons for optimum target thicknesses from simulations (dots) and theory
(lines).

It is clearly seen that the theory correctly reproduces the dependence of the maximum proton energy on laser intensity.
Since our theory considers single-ion-species (H) foil, the curves for analytical solutions in Fig \ref{pic_e_a1} slightly overestimate simulation results (up to 30\% for $\tau$=30 fs). This is expected because carbon ions from CH$_2$ foil are also accelerated thus making proton energy lower.
Note that the apparent and very good agreement between theory and simulations for the long pulses (150 fs) is due to another limitation of the theoretical model that disregards the effect of the transversal (radial) electron current from the periphery of the focal spot to its centre.  Such a flow of particles is able to provide more accelerated electrons on the rear side of the target for the long enough laser pulses. For $\tau>d/\upsilon_{\perp}$, this effect may be significant. The velocity $\upsilon_{\perp}$ changes from $0$ to $c$ and the typical time of transversal electron  motion, $\sim d/c$, will be in the order of a few tens of femtoseconds. Notwithstanding these approximations, Fig \ref{pic_e_a1} demonstrates that overall, our simple theory correctly describes dependence  of the maximum ion energy on laser and plasma parameters.

In summary, we have studied proton acceleration from ultra-thin targets
with optimal thicknesses. For the first time, the absorption of laser light by
semi-transparent plasma has been described. This permits an understanding of laser intensity dependence of maximum ion energy, $\e \sim E_l^{0.7},$ for a wide range of laser energies (from 0.03
J to 300 J). A simple analytic theory has been proposed for the wide range of laser
parameters. It agrees well with our simulation results.
Both analytical and numerical predictions are quite optimistic about the ion energy gain even for
laser intensities below the range required for radiation pressure (light
sail) regime. In general, experiments show that ion energy increases slower with laser intensity as compared to scaling predictions. We believe that the systematic experimental study with targets of optimum thicknesses and high contrast laser pulses should confirm the theoretical predictions of our paper.

This work was supported by the Russian Foundation for Basic Research
(Grants Nos. 14-02-31407-mol\_a, 13-02-00426-a, 12-02-00231-a). Research of WR was partially supported by the Natural Sciences and Engineering Research Council of Canada.

\end{document}